\documentclass[aip,twocolumn]{revtex4}
\usepackage{epsfig,graphics,amssymb,amsmath,subeqnarray}
\usepackage{color}

\def\f{{\mathbf{f}}}
\def\t{{\mathbf{t}}}

\def\u{{\mathbf{u}}}

\def\U{{\mathbf{U}}}

\def\R{{\mathbf{R}}}
\def\r{{\mathbf{r}}} 

\def\br{{\bar{\mathbf{r}}}}

\def\varepsilon{\epsilon}

\begin{document}

\title{Extensibility enables locomotion under isotropic drag}

\author{On Shun Pak and Eric Lauga\footnote{Corresponding author. Email: elauga@ucsd.edu}}
\affiliation{
Department of Mechanical and Aerospace Engineering, 
University of California San Diego,
9500 Gilman Drive, La Jolla CA 92093-0411, USA.}

\begin{abstract}

Anisotropic viscous drag  is usually believed to be a requirement for the  low Reynolds number locomotion of slender bodies such as flagella and cilia. Here we  show that  locomotion under isotropic drag is possible for extensible slender bodies. After general considerations, a two-ring swimmer and a model dinoflagellate flagellum are studied analytically to illustrate how extensibility can be exploited for self-propulsion without drag anisotropy.   This new degree of freedom could be useful for some complex swimmer geometries and locomotion in complex fluid environments where drag anisotropy is weak or even absent.

\end{abstract}

\maketitle

Due to the absence of inertial forces, low-Reynolds number locomotion is subject to interesting mathematical and physical constraints  \cite{brennen,fauci3,lauga2}. In particular, locomotion by time-reversible strokes is ruled out by Purcell's scallop theorem \cite{purcell77,lauga11}.  To escape these constraints, microorganisms swim by either propagating deformation waves along slender appendages, termed flagella, or rotating them.  Anisotropic viscous drag is believed to be the fundamental property enabling drag-based propulsion of slender filaments \cite{hancock,becker,lauga2}. It is a classical result that for a slender filament moving in an unbounded Newtonian fluid, the Stokes drag is  almost twice  when moving perpendicular than parallel to its axis \cite{happel,lauga2}. This  drag anisotropy allows propulsive forces to be created perpendicularly to the deformation of the filament. Under isotropic drag, it is generally accepted that locomotion of this kind would be impossible \cite{hancock,gray55,becker,camalet00,yu06}. 

Unlike in Newtonian flows, drag laws in more complex media, and their consequences on locomotion, remain largely unexplored. Theoretical studies, via Brinkman models, suggest that porosity enhances drag anisotropy \cite{leshansky, jung}, explaining, e.g., the increase in propulsion speed of \textit{C. elegans} in a granular medium. Recent experiments also measured and characterized granular drag in beds of glass beads and granular media \cite{soller,hill,ding,gravish}, which have been applied to study locomotion in sand \cite{maladen}. {Besides the fluid medium, the geometry of a swimming body also plays a role in the drag law. Some flagella, such as those of \textit{Ochromonas}, possess rigid projections termed mastigonemes, protruding into the fluid \cite{brennen}. In these geometries  the viscous drag in the longitudinal direction of the flagellum is increased, possibly resulting in a more isotropic drag.} In situations where drag anisotropy is weak or even absent, what are the alternative mechanisms, if any, offered by physics to achieve locomotion? In this letter, we point out  a new degree of freedom enabling inertialess swimming, namely  extensibility.  Using a general derivation and two simple geometrical models, we demonstrate that the periodic stretching and contraction of a filament allow self-propulsion even under isotropic drag. 

We start by considering the  general calculation of Becker \textit{et al.}~\cite{becker}  showing that drag anisotropy is required  for the propulsion of inextensible swimmers. Here we revisit their derivation by relaxing the inextensibility condition.  Consider a filamentous swimmer of total length $L(t)$, and denote 
by $\r (s,t)$  the instantaneous position  of material points along the filament, where $s$ is the arclength. The time rate of change of the average swimmer position, $\br(t)$,  is given by
\begin{align}
\frac{d \br}{dt}& = \frac{d}{dt} \left[ \frac{1}{L(t)} \int_0^{L(t)} \r(s,t)ds \right]\\
&= \frac{{\dot{L}}}{L} \left[ \r(L,t)-\frac{1}{L} \int_0^{L}\r ds \right]+\frac{1}{L}\int_0^{L} \frac{\partial  \r}{\partial t} ds,
\label{eqn:second}
\end{align}
where we have denoted ${\dot{L}}\equiv dL/dt$. Under isotropic drag, we have ${\partial  \r}/{\partial t} ={\bf u} \propto \f $, where $\f$ is the local drag  force per unit length, and thus $\int_0^{L} {\partial  \r} / {\partial  t} ds \propto \int_0^{L} \f ds=\bf 0$ for force-free swimming. We therefore find that the second term in Eq.~\eqref{eqn:second} disappears, and thus
\begin{align}
\frac{d \br}{dt}= \frac{{\dot{L}}}{L} \left[ \r(L,t)-\br \right].
\end{align}
For an inextensible swimmer, ${\dot{L}}(t)=0$, leading to $d \br/dt =0$: no net propulsion is possible under isotropic drag \cite{becker}.  As a difference, for an extensible swimmer (${\dot{L}}\neq0$) no general conclusion can be drawn, suggesting the relaxation of the drag anisotropy requirement  when extensibility is permitted. We demonstrate below how this additional degree of freedom can be exploited for self-propulsion by considering two simple examples.

\begin{figure*}[t]
\begin{center}
\includegraphics[width=0.95\textwidth]{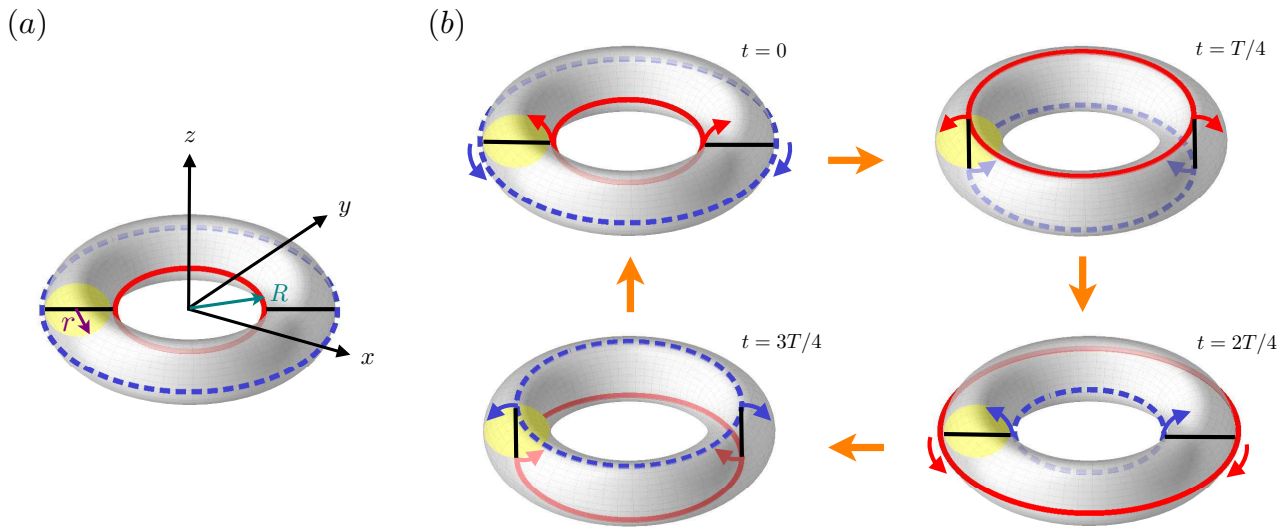} 
\end{center}
\caption{\label{fig:setup}(a) Notation for the extensible two-ring swimmer. 
(b) Schematic illustration of its cyclic deformation over one period, $T$ (see text for details).
\label{fig:motion}}
\end{figure*}

As the first example, consider the motion of two extensible slender rings (red solid and blue dashed circles in Fig.~\ref{fig:setup}) connected by two rotating rigid rods (each of length 2$r$). As a result of the rod rotation, each ring expands and contracts periodically, tracing in time  the surface of a torus of inner radius $R$ and a circular cross-section of radius $r$ (see Fig.~\ref{fig:setup}a).  
In Cartesian coordinates, the motion of each ring $\r (t)= [x(t),y(t),z(t)]$  can be described as
\begin{align}
x(t) &= \left\{R+ r \left[ 1- \cos (\omega t +\phi_0)\right] \right\} \cos \theta,\\
y(t) &= \left\{R+ r \left[ 1- \cos (\omega t +\phi_0)\right] \right\} \sin \theta,\\
z(t) &= r \sin(\omega t+\phi_0),
\end{align}
where, $\theta \in [0, 2 \pi]$, $\omega = 2\pi/T$ is the angular frequency, $T$ is the period of the motion, and $\phi_0$ is the phase of the motion. We non-dimensionalize lengths by the inner radius of the torus $R$, time by $1/\omega$, and the dimensionless deformation kinematics of the two rings, $\tilde{\r}_{1,2}(\tilde{t}) = [\tilde{x}_{1,2}(\tilde{t}), \tilde{y}_{1,2}(\tilde{t}), \tilde{z}_{1,2}(\tilde{t})]$, are given by
\begin{align}
\tilde{x}_{1,2} (\tilde{t})&= \left\{1+ \tilde{r} \left[ 1- \cos (\tilde{t} +\phi_{1,2})\right] \right\} \cos \theta,\\
\tilde{y}_{1,2} (t)&= \left\{1+ \tilde{r} \left[ 1- \cos (\tilde{t} +\phi_{1,2})\right] \right\} \sin \theta,\\
\tilde{z}_{1,2}(\tilde{t}) &= \tilde{r} \sin(\tilde{t}+\phi_{1,2}),
\end{align}
where  we assign $\phi_1 = 0 $ (red solid ring) and $\phi_2=\pi$ (blue dashed ring).  All tilde variables in this letter are dimensionless.

We illustrate the cyclic deformation of this swimmer over  one period in Fig.~\ref{fig:motion}b. Hydrodynamically, material is being created/destroyed when a ring extends/contracts, introducing an additional mechanism for varying the hydrodynamic drag. The non-zero translational velocity expected to arise is a direct result of the imbalance of vertical viscous force due to the difference in the total arc-length of the two rings.  Since the vertical motion and variation of the circumference of the two rings are out-of-phase, the {vertical} velocity of the ring with a larger circumference  always points in the same direction (in the case considered here, in the  $-z$ direction). Therefore, we expect a net unidirectional  vertical force, and {hence} swimming, in this direction. Note that when the two rings have, instantaneously, exactly the same diameters (at $\tilde{t} = (2 {n}+1)\pi/2$, where ${n}=0,1,2,... \ $; see the upper right and lower left panels in Fig.~\ref{fig:motion}b), the viscous forces acting on the  rings balance, and no swimming is produced. Note also that if there was  only one ring, say $\tilde{\mathbf{r}}_1$, the vertical motion of the ring,  would  be exactly canceled by  translational swimming velocity at the same speed, resulting in no apparent motion.

Quantitatively, we apply a general local drag theory to this two-ring swimmer. We neglect the hydrodynamic impact of the rods connecting the rings.  The local viscous force density acting on the filament is then given by $\f = -[\xi_{\parallel} \t \t + \xi_{\perp} (1-\t \t)] \cdot \u$, where $\t$ is the local tangent vector and $\u$ is the local velocity of the filament. The drag coefficients $\xi_{\parallel}$ and $\xi_{\perp}$ characterize the motion of a slender rod parallel and perpendicular to its axis respectively. Their specific values depend on the geometry of the rod and properties of the fluid medium. For a slightly distorted slender filament in an unbounded Newtonian fluid, Gray and Hancock \cite{hancock, gray55} derived explicit analytical expressions for the drag coefficients, which were later improved by Lighthill \cite{lighthill}. For more complex environments, these drag laws remain mostly unknown. Under isotropic drag, we have $\xi_{\perp} = \xi_{\parallel}$  \cite{shelley00,goldstein95,bourdieu,goldstein98}. Here, we  keep their values  general in the calculations, and show that the final swimming speed of the two-ring swimmer is independent of these coefficients. The swimmer hence works equally well under any local drag law,  including isotropic drag.

\begin{figure}[t]
\begin{center}
\includegraphics[width=0.4\textwidth]{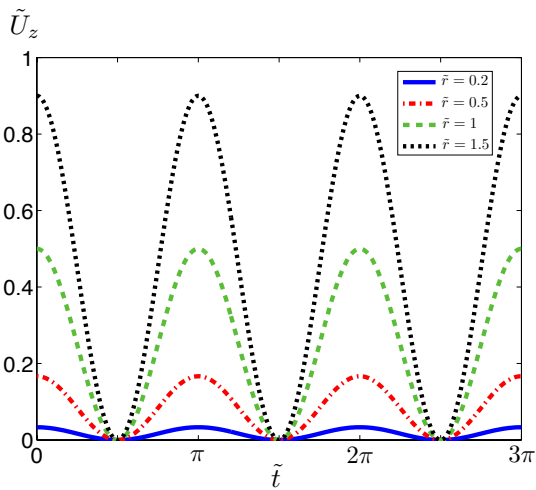}
\end{center}
\caption{\label{fig:Uz}Time-variation of the dimensionless swimming velocity of the extensible two-ring swimmer, $\tilde{U}_z$,  for different values of $\tilde{r}$.}
\end{figure} 

The local velocity along a ring,  $\u = \partial \r/ \partial t+\U+\mathbf{\Omega} \times \r$, is composed of two parts: the prescribed deformation velocity, $\partial \r/ \partial  t$, and the unknown swimming velocities $\U+\mathbf{\Omega} \times \r$ to be determined. At low Reynolds numbers, the total force and total torque on a swimmer have to vanish. Hence, we have $\sum_{i=1}^{2} \int_0^{L_i(t)} \f_i ds_i=\sum_{i=1}^{2} \int_0^{L_i(t)} \boldsymbol\tau_i ds_i = \bf 0$, where $\boldsymbol\tau = \r \times \f$ is the local viscous torque density, leading to six equations to determine the unknown swimming kinematics $(\U,\mathbf{\Omega})$.  By symmetry, we have $\mathbf{\Omega}=\bf 0$, and $U_z$ is the only  non-zero  translational velocity component. After some algebra we obtain the dimensionless swimming velocity
\begin{align}
\tilde{U}_z = \left(\frac{\tilde{r}^2}{1+\tilde{r}} \right) \cos^2 \tilde{t}.
\end{align}
We display the variation of the swimming velocity as a function of time for different values of $\tilde{r}$ in Fig.~\ref{fig:Uz}. We obtain  unidirectional swimming ($\tilde{U}_z \ge 0$) at a time-averaged speed independent of the value of the drag coefficients, $\langle \tilde{U}_z\rangle =\tilde{r}^2/2(1+\tilde{r})$: this extensible two-ring swimmer can therefore self-propel under  isotropic drag.

\begin{figure}[t]
\begin{center}
\includegraphics[width=0.42\textwidth]{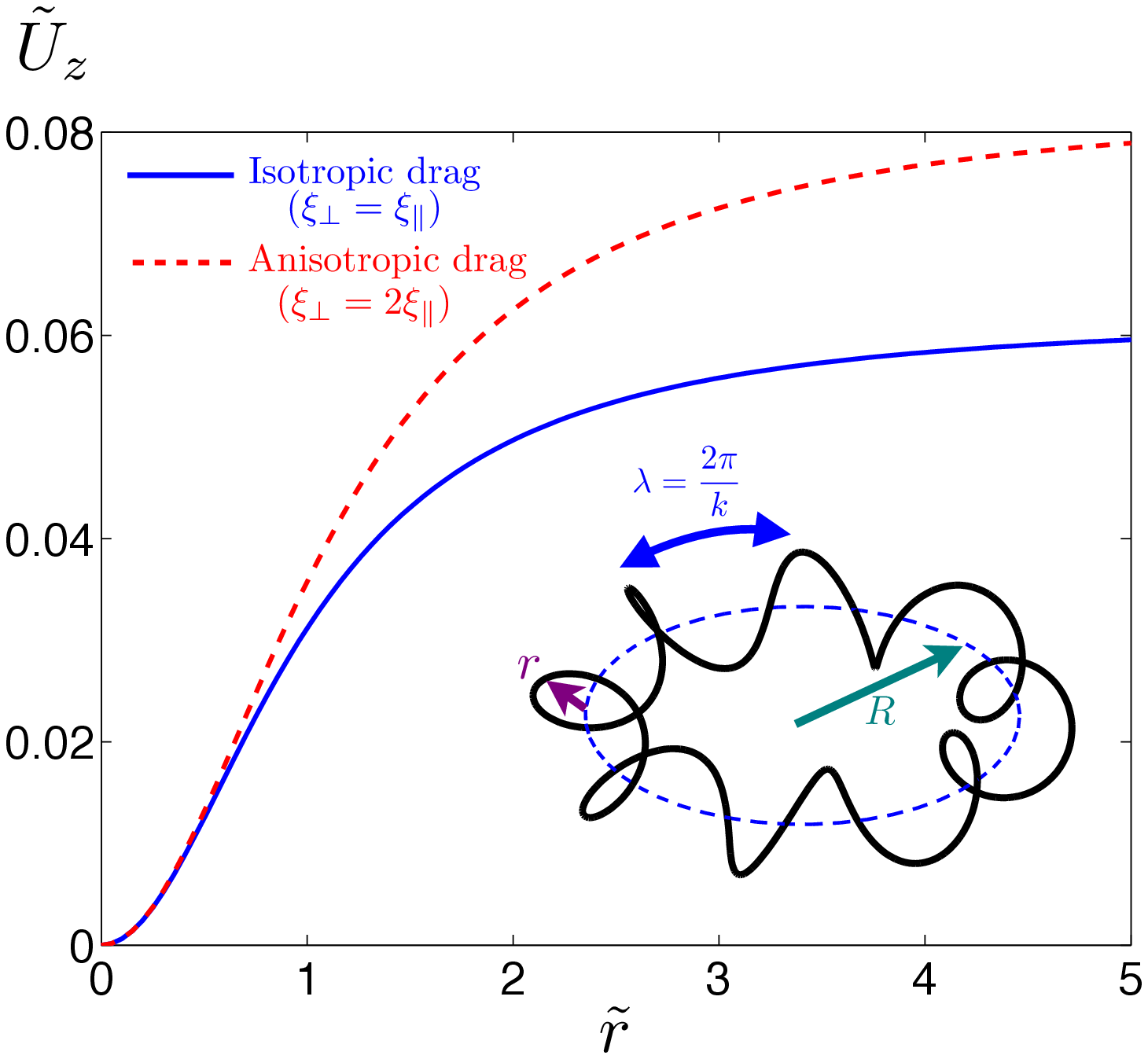}
\end{center}
\caption{\label{fig:toroidal} Dimensionless translational velocity $\tilde{U}_z$ as a function of the dimensionless toroidal wave amplitude, $\tilde{r} = rk$, under isotropic ($\xi_{\perp} = \xi_{\parallel}$, blue solid line) and anisotropic ($\xi_{\perp} = 2\xi_{\parallel}$, red dashed line) drag, for the case of $\tilde{R}=8$. Inset: notation for a toroidal helix swimmer (see text for details).}
\end{figure}

This  idea can be extended to more complicated geometries consisting of a curved structure built upon another curved structure, for instance, toroidal helices (a helix built upon a circle) and superhelices (a helix built upon another helix). A toroidal helix has been studied as an idealized geometrical model for dinoflagellates \cite{miyasaka, nguyen}. A dinoflagellate can be propelled by propagating a toroidal helical wave, where the kinematics of such a wave implicitly assumes extensibility due to the intrinsic length differences in the geometry. Consider a toroidal helical wave with amplitude $r$, wave-number $k=2\pi/\lambda$ and angular frequency $\omega$, propagating along a circle of radius $R$ (see Fig.~\ref{fig:toroidal} inset). When lengths are non-dimensionlized by $1/k$ and times by $1/\omega$, the dimensionless kinematics $\tilde{\R} = [\tilde{x}', \tilde{y}', \tilde{z}']$ can be expressed as
\begin{align}
\tilde{x}'(\tilde{s}',\tilde{t}) &= \tilde{R} \cos(\tilde{s}'/\tilde{R}) + \tilde{r} \cos(\tilde{s}'-\tilde{t}) \cos(\tilde{s}'/\tilde{R}),\\
\tilde{y}'(\tilde{s}',\tilde{t}) &= \tilde{R} \sin(\tilde{s}'/\tilde{R})+\tilde{r} \cos(\tilde{s}'-\tilde{t}) \sin (\tilde{s}'/\tilde{R}),\\
\tilde{z}'(\tilde{s}',\tilde{t}) &= \pm \tilde{r} \sin(\tilde{s}'-\tilde{t}),
\end{align}
where  $\tilde{s}' \in [0, 2\pi \tilde{R}]$ parametrizes the toroidal helix, and where all tilde parameters are dimensionless. Note that the dimensionless radius $\tilde{R}$ takes only integer values for a closed toroidal helix. The $\pm$ sign represents different chirality of the helix. We apply the same local drag model as above. In order to make analytical progress, we consider the small-amplitude limit $\tilde{r} \ll 1$ and perform asymptotic expansions in powers of $\tilde{r}$. Swimming occurs at $O(\tilde{r}^2)$ and takes the dimensionless form
\begin{align}
\tilde{\U} = \left[0, 0, \pm \frac{\tilde{r}^2}{2\tilde{R}} \right], \quad 
\tilde{\mathbf{\Omega}} = \left[0, 0, \frac{1-\xi_{\perp}/\xi_{\parallel}}{\tilde{R}}\tilde{r}^2\right].
\end{align}

While the rotational velocity $\tilde{\mathbf{\Omega}}$ vanishes under isotropic drag ($\xi_{\perp}/\xi_{\parallel}=1$), the translational velocity $\tilde{\U}$ along the torus axis is independent of the drag coefficients.  This is due to the intrinsic difference in length (and hence drag) embedded in the curved geometry of toroidal helices. The toroidal helix has shorter lengths on the sides closer to the center of the circle, creating an overall imbalance of hydrodynamic drag. {It is interesting to compare two physical limits, namely when there is no drag anisotropy ($\xi_{\perp} = \xi_{\parallel}$) and when there is maximum drag anisotropy for an asymptotically slender filament in a Newtonian flow ($\xi_{\perp} = 2\xi_{\parallel}$). Using these values in the general local drag theory, we find that the axial propulsion velocity in these two limits are indistinguishable for small $\tilde{r}$ (Fig.~\ref{fig:toroidal}), as predicted by the asymptotic analysis. As $\tilde{r}$ increases, the drag anisotropy increasingly enhances propulsion (up to $\sim 30\%$ when $\tilde{r}=5$) but extensibility alone, under isotropic  drag, still enables swimming with significant magnitudes.} A similar example that could exploit extensibility to produce swimming under isotropic drag are superhelical waves, which serve as a geometrical model for insect spermatozoa \cite{werner08}.

In conclusion, in this letter we have revisited the requirement of drag anisotropy to achieve self-propulsion at zero Reynolds number. We demonstrated explicitly, via two simple swimmers, that extensibility provides a mechanism for swimming under isotropic drag, which might be relevant for some complex swimmer geometries and motion in porous or other complex media where drag anisotropy is weak or absent. As a practical side-note, the two-ring swimmer described above could be actuated experimentally by external rotating magnetic fields if ferromagnetic materials, such as metal nanowires \cite{bentley,keshoju}, were used for the phantom rods.

\section*{Acknowledgments}
We thank Saverio Spagnolie and Gwynn Elfring for useful discussions. Funding by the National Science Foundation (Grant No. CBET-0746285 to E.~L.), and the Croucher Foundation (through a scholarship to O.~S.~P.) is gratefully acknowledged.

\end{document}